\documentclass[10pt,a4paper]{article}

\usepackage{amsmath}
\usepackage{amssymb}
\usepackage{amsfonts}
\usepackage{pstricks}
\usepackage{color}
\usepackage{subfigmat}
\usepackage[dvipdfmx]{graphicx}
\usepackage{latexsym}
\usepackage{cite}
\usepackage{comment}
\usepackage{fancyhdr}
\usepackage{authblk}
\usepackage[top=30truemm,bottom=30truemm,left=25truemm,right=25truemm]{geometry}

\title{
\Large{\textbf{Charged Lepton Flavor Violation in the Semi-Constrained NMSSM
with Right-Handed Neutrinos}}}

\author[1]{\large{Keisuke Nakamura}\thanks{
\texttt{knakamura@tuhep.phys.tohoku.ac.jp}}}
\author[2]{\large{Daisuke Nomura}\thanks{
\texttt{dnomura@yukawa.kyoto-u.ac.jp}}}

\affil[1]{\normalsize{Department of Physics, Tohoku University,
 Sendai 980-8578, Japan}}
\affil[2]{\normalsize{Yukawa Institute for Theoretical Physics,
 Kyoto University, Kyoto 606-8502, Japan}}

\date{}

  \makeatletter
    
    \@addtoreset{equation}{section}
  \makeatother

\def\hH{{\hat H}}
\def\hQ{{\hat Q}}
\def\hU{{\hat U}}
\def\hD{{\hat D}}
\def\hL{{\hat L}}
\def\hN{{\hat N}}
\def\hE{{\hat E}}
\def\hS{{\hat S}}

\def\vu{v_{u}}
\def\vd{v_{d}}

\def\mueff{\mu_{{\rm eff}}}
\def\Beff{B_{{\rm eff}}}

\def\mhu{m^{2}_{H_{u}}}	
\def\mhd{m^{2}_{H_{d}}}
\def\ms{m^{2}_{S}}

\def\gi{g_{1}}
\def\gii{g_{2}}

\def\mz{M_{Z}}

\def\Tr{{\rm Tr}}

\def\tm2{\widetilde{m}^{2}}

\def\m2{m^{2}}

\def\>{\rangle}
\def\<{\langle}

\def\m2{m^{2}}

\begin{document}
\maketitle

\thispagestyle{fancy}
\rhead{{TU-989, YITP-14-108}}

\begin{abstract}
 \noindent
We study the $\mu \to e \gamma$ decay in the 
$\mathbb{Z}_{3}$-invariant next-to-minimal supersymmetric (SUSY)
Standard Model (NMSSM) with superheavy right-handed neutrinos.
We assume that the soft SUSY breaking parameters are 
generated at the GUT scale, not universally as in the
minimal supergravity scenario but in such a way that
those soft parameters which are specific to the NMSSM
can differ from the soft parameters which involve
only the MSSM fields while keeping the universality at the GUT scale
within the soft parameters for the MSSM and right-handed neutrino
fields.  We call this type of boundary conditions 
``semi-constrained''.
In this model, the lepton-flavor-violating off-diagonal
elements of the slepton mass matrix are induced
by radiative corrections from the neutrino Yukawa couplings,
just like as in the MSSM extended with the right-handed
neutrinos, and these off-diagonal elements induce sizable
rates of $\mu \to e \gamma$ depending on the parameter
space.  Since this model has more free parameters than
the MSSM, the parameter region favored from the Higgs
boson mass can slightly differ from that in the MSSM.
We show that there is a parameter region in which
the $\mu \to e \gamma$ decay can be
observable in the near future even if
the SUSY mass scale is about 4 TeV.

\end{abstract}

\section{\Large{Introduction}}
It is now clear that the lepton flavor number is not a conserved
quantity because of experimental observations of neutrino 
oscillations~\cite{Agashe:2014kda}.
In the minimal extensions of the Standard Model (SM) with
the Majorana neutrino
mass terms, the branching ratios for charged lepton-flavor violating
(LFV) processes are extremely small since they are suppressed
by at least a factor of $m_\nu^2/m_W^2$, which makes it very difficult
for near-future experiments to detect LFV signals. 
On the other hand, in more general extensions of the SM, which 
are motivated by several reasons, it is known that sizable
LFV rates are predicted depending on parameter region.
If LFV processes are discovered, it directly means an
indirect signature of physics beyond the SM (BSM).  
Recently,  the MEG experiment reported a new upper limit
of ${\rm Br}(\mu\to e\gamma) < 5.7 \times 10^{-13}$~\cite{Adam:2013mnn}.
This already gives a strong constraint on models beyond the SM,
and hence it is very important to keep updating these upper
bounds on the LFV processes.

Supersymmetry (SUSY) is still a promising candidate for
physics beyond the SM~\cite{Martin:1997ns}.  Lots of effort
has been devoted to the discovery of SUSY at the LHC, but only
in vain so far.  The most studied model of SUSY is the minimal
SUSY SM (MSSM).  Even in the framework of the MSSM, there are 
some unsolved problems such as the $\mu$ problem.  
Next-to-the MSSM (NMSSM) is an extension of the MSSM
with a SM-singlet Higgs chiral superfield $\hat{S}$.  The NMSSM
could give a hint to solve the $\mu$ problem since in this
model the $\mu$ term is induced by the vacuum expectation value
(VEV) of the scalar component $S$ of $\hat{S}$.
In this sense the NMSSM is a natural extension of the MSSM.

One of the difficulties in the MSSM is the Higgs boson mass.
In the MSSM, the tree-level lightest Higgs boson mass is
bounded from above as,
\begin{align}
 m^{2}_{h,{\rm MSSM}} \bigg|_{\rm tree}
 < \mz^{2} \cos^2{2\beta}~,
\end{align}
and has to
rely on large radiative corrections to reproduce the observed
Higgs boson mass of 126 GeV~\cite{Agashe:2014kda}.
The main contribution to the radiative corrections comes
from the top Yukawa coupling~\cite{Okada:1990vk, Ellis:1990nz,
Haber:1990aw}, and to maximize this effect
one needs a top-squark mass much larger than the top-quark
mass.  In the NMSSM, the lightest Higgs boson mass 
reads~\cite{Ellwanger:2009dp}:
\begin{align}
m^{2}_{h,{\rm NMSSM}}\approx\mz^{2}\cos^2{2\beta}
+\lambda^{2}v^{2}\sin^2{2\beta}
+\Delta m^{2}_{h,{\rm 1Loop}} ~,
	\label{eq:apx_higgs}
\end{align}
where $v\sim 174$ GeV. 
As is seen from this equation, the contribution from the new
parameter $\lambda$, which is the coupling among
the new singlet $S$ and the MSSM Higgs doublets $H_u$ and $H_d$,
makes the tree-level Higgs boson mass larger, in particular
for small $\tan\beta$.  We have to note that 
the mixings between $S$ and the MSSM Higgs doublets
can make a negative contribution to the lightest Higgs boson mass,
and the NMSSM does not always predict a larger Higgs boson mass.
We will discuss this issue in details later in this paper.

There are more than one-hundred free parameters in the MSSM.
Usually, we assume an underlying scenario for SUSY breaking,
and it allows us to reduce the number of free parameters.
In this paper we assume the minimal supergravity (mSUGRA)-like
boundary conditions
that the SUSY breaking parameters $m_0, M_{1/2}, A_0$ are universal
at the GUT scale.  The parameters at the SUSY scale are obtained
by evolving these parameters according to the renormalization
group equations (RGE).  These mSUGRA-like boundary conditions
are very effective for avoiding constraints from the SUSY-induced
flavor changing neutral current (FCNC) processes.  This is also
true for the charged LFV processes, and in the mSUGRA, also
known as the constrained MSSM (cMSSM), there are essentially
no charged LFV.  This is similar in the case of the constrained
NMSSM.

The neutrino masses are exactly zero in the framework of
the SM, which clearly needs modifications in view of the
observation of neutrino oscillations.  One of the most natural
mechanisms to explain the tiny neutrino masses is the (type-I)
seesaw mechanism~\cite{Minkowski:1977sc, Yanagida:1979as,
GellMann:1980vs}, which we consider in this paper.  The
extension of the original seesaw mechanism to SUSY models
is straightforward.  In the MSSM extended with the right-handed
neutrinos $\nu_R$, which we call the MSSM $+~\nu_R$ model,
even if one assumes the mSUGRA-like boundary
conditions at the GUT scale,
off-diagonal elements in the slepton  mass matrices are induced
via radiative corrections from the neutrino Yukawa couplings,
which can predict sizable rates for the LFV processes
like $\mu \to e \gamma$.  This mechanism also works in the
NMSSM extended with the right-handed neutrinos, which
we call the NMSSM $+~\nu_R$ model and which we
consider in this paper, but since there are more free parameters
than in the case of the MSSM $+~\nu_R$ model,
the predicted LFV rates can slightly differ from those in the
MSSM $+~\nu_R$ model in the parameter region favored from
the Higgs boson mass.

The contents of this paper are as follows.  In Section 2,
we introduce the model we work with, and in Section 3 we explain the
origin of the LFV (off-diagonal) elements of the slepton 
mass matrices.  In Section 4, we discuss constraints
on the parameters of the model.  We introduce the results
of numerical calculations in Section 5, and in Section 6
we summarize this paper.

\section{\Large{Model}}
\subsection{$\mathbb{Z}_{3}$-invariant NMSSM}

The NMSSM is an extension of the MSSM, and it has an
extra Higgs chiral superfield $\hS$ which is singlet under the
SM gauge group.  In the $\mathbb{Z}_{3}$-invariant
NMSSM~\cite{Ellwanger:2009dp},
the $\mu$ term $\mu \hH_u\cdot \hH_d$ in the superpotential
of the MSSM is replaced by the term
$\lambda \hS\hH_u\cdot \hH_d$, and the $\mu$-parameter 
is determined from the singlet VEV $s$ as $\mu_{{\rm eff}}=\lambda s$.
Namely, the superpotential of the $\mathbb{Z}_{3}$-invariant
NMSSM is given as 
\begin{align}
    W_{{\rm NMSSM}} = W_{{\rm MSSM}}\big|_{\mu=0}
        + \lambda\hS\hH_{u}\cdot\hH_{d}
        + \frac{1}{3}\kappa\hS^{3} ~,
\end{align}
where the dot in the term $\lambda \hS \hH_{u}\cdot\hH_{d}$ represents 
the $SU(2)$-invariant product of two $SU(2)$ doublets, and the hats on the
fields stand for the superfields corresponding to the
fields\footnote{If we are to consider more general NMSSM, not the
$\mathbb{Z}_{3}$-invariant NMSSM, we will have more terms
in the superpotential, such as the terms linear and quadratic in
$\hat{S}$~\cite{Ellwanger:2009dp}, and also the soft SUSY breaking
terms associated with them.  To reduce the number of free
parameters, in this paper we only consider the $\mathbb{Z}_{3}$-invariant
version of the NMSSM+$\nu_R$ model.}.
We assume that the $R$-parity  is conserved, and assign
the even $R$-parity to $\hS$.
The soft SUSY breaking terms are
\begin{align}
V_{{\rm soft}} &= V_{{\rm MSSM}}\big|_{\mu=B=0}
   + \left(\lambda A_{\lambda}S H_{u}\cdot H_{d} 
 + \frac{1}{3}\kappa A_{\kappa}S^{3} + {\rm h.~c.} \right)
 +m^{2}_{S}|S|^{2} ~.
\end{align}

In the case of the constrained NMSSM, the gaugino masses,
sfermion soft SUSY breaking masses, and the $A$-parameters
take the values which are ``universal'' at the GUT scale,
similarly to the case of the cMSSM:
\begin{align}
    M_\alpha &= M_{1/2} \label{eq:tmp2.3}\\
 m^{2}_{H_u} &= m^{2}_{H_d}= m^2_0 ~,\\
 (m^{2}_{f})_{ij} &= m^{2}_{0} \delta_{ij} ~~~~
 (f=Q,U,D,L,E)~,\\
    (A_f)_{ij} &= A_{0} \delta_{ij} ~~~~~ (f=U,D,E)~,\label{eq:tmp2.6}
\end{align}
where $\alpha$ ($\alpha=1,\ldots,3$) labels the gauge groups of the
SM, and $i$ and $j$ are the indices for generations, $i,j= 1,\ldots,3$.
As for the parameters $A_{\lambda},\ A_{\kappa}$ and $m^{2}_{S}$
which are specific to the NMSSM, we
assume that the values of $A_\lambda$ and $A_\kappa$ at the GUT scale
are not necessarily equal to $A_0$,
and that $m^{2}_{S}$ at the GUT scale can be different from $m_0^2$. 
We call the NMSSM with this class of boundary
conditions the semi-constrained NMSSM.

\subsection{$\mathbb{Z}_{3}$-invariant NMSSM
extended with right-handed neutrinos}

In this paper we take the simplest extension of the 
$\mathbb{Z}_{3}$-invariant NMSSM with the right-handed
neutrinos, in which the (type-I) seesaw
mechanism~\cite{Minkowski:1977sc,Yanagida:1979as,GellMann:1980vs}
is at work.  The superpotential is given by
\begin{align}
   W = W_{{\rm NMSSM}} 
     + (Y_{N})_{ji}\hH_{u}\cdot\hL_{i}\hN^{c}_{j}
     + \frac{1}{2}(M_N)_{ij}\hN^{c}_{i}\hN^{c}_{j}~,
\end{align}
where the $\mathbb{Z}_{3}$-charges are assigned as in 
Table~\ref{tbl:z3charge}\cite{Gogoladze:2008wz}.
This charge assignment excludes the term
$(\lambda_{\nu})_{ij}\hS\hN^c_i\cdot\hN^c_j$
from the superpotential\footnote{
It is possible to derive the (left-handed) neutrino masses
via the type-I seesaw mechanism 
from the Majorana masses which emerge from term
$(\lambda_{\nu})_{ij}\hS\hN^c_i\cdot\hN^c_j$ after replacing $S$ with its VEV.
In this case, since the singlet VEV $\langle S \rangle$ is at most
${\cal O}(1-100{\rm ~TeV})$, the Majorana masses must be about
the same order, which forces us to assume a very small neutrino Yukawa
coupling $(Y_N)$ in order to explain the tiny neutrino masses.
This makes the LFV rates extremely small
and hence we do not consider this scenario in this paper.
We should also note that in some of extended seesaw schemes such
as the inverse seesaw mechanism~\cite{InverseSeesaw}, it is
possible to have the neutrino Yukawa couplings of ${\cal O}(0.01)$
and the right-handed neutrinos of the mass of the order of the EW
scale simultaneously.  It is known in literature that in the
SUSY inverse seesaw models the predicted LFV rates can be
sizable~\cite{DeppischValle}.}.
\begin{table}[t]
\begin{center}
\begin{tabular}{|c|c|c|c|c|c|c|c|c|c|} \hline
                  & $\hQ$     & $\hU^{c}$ & $\hD^{c}$ & $\hL$ & $\hE^{c}$
   & $\hN^{c}$    & $\hH_{u}$ & $\hH_{d}$ & $\hS$ \\ \hline \hline
 $\mathbb{Z}_{3}$ charges & $\omega^{2}$ & 1 & 1 & $\omega^{2}$ & 1 
   & 1 & $\omega$ & $\omega$ & $\omega$\\ \hline
\end{tabular}
\caption{The $\mathbb{Z}_{3}$-charge assignment in the
$\mathbb{Z}_{3}$-invariant NMSSM extended with the right-handed
neutrinos.  In the table, $\omega \equiv e^{2\pi i/3}$.}
\label{tbl:z3charge}
\end{center}
\end{table}

The neutrino masses in this model is 
\begin{align}
    (m_{\nu})_{ij}
      &=v^{2}\sin^{2}{\beta}(Y_N)_{ki}((M_N)^{-1})_{kl}(Y_N)_{lj} \nonumber\\
      &=(U_{{\rm MNS}}^{\top})_{ik}m_{\nu,k}(U_{{\rm MNS}})_{kj} ~,
\label{eq:tmp_2.8}
\end{align}
where $U_{{\rm MNS}}$ is the MNS matrix~\cite{Maki:1962mu}
and $m_{\nu, k}$ $(k=1,\ldots, 3)$ are the eigenvalues of the
left-handed neutrino mass matrix $(m_{\nu})_{ij}$.
In the standard representation of the PDG, the matrix reads:
\begin{align}
U_{{\rm MNS}}&=
	\left(
	\begin{array}{ccc}
	c_{12}c_{13}&s_{12}c_{13}&s_{13}e^{-i\delta}\\
	-s_{12}c_{23}-c_{12}s_{13}s_{23}e^{i\delta}&c_{12}c_{23}-s_{12}s_{13}s_{23}e^{i\delta}&c_{13}s_{23}\\
	s_{12}s_{23}-c_{12}s_{13}c_{23}e^{i\delta}&c_{12}s_{23}-s_{12}s_{13}c_{23}e^{i\delta}&c_{13}c_{23}
	\end{array}
 \right)
 \nonumber\\
 &\quad\times{\rm diag}(1,e^{i\alpha_{21}/2},e^{i\alpha_{31}/2})~,
\end{align}
where $c_{ij}=\cos{\theta_{ij}},s_{ij}=\sin{\theta_{ij}}$.  
The mixing angles
$\theta_{ij}$ $(i,j=1,\ldots,3, i<j)$ describe the mixing between the
mass eigenstates $\nu_{i}$ and $\nu_{j}$, and
the factors $\delta,\alpha_{21},\alpha_{31}$ are 
complex phases, and represent
the Dirac phase and the two Majorana phases, respectively.
According to the latest data~\cite{Agashe:2014kda}
the values of the angles are:
\begin{align}
\sin^{2}{2\theta_{12}}&=0.846\pm0.021 \label{eq:tmp_2.10}\\
\sin^{2}{2\theta_{23}}&=
\begin{cases}
0.999^{+0.001}_{-0.018}\ ({\rm normal\ mass\ hierarchy})\\
1.000^{+0.000}_{-0.017}\ ({\rm inverted\ mass\ hierarchy})
\end{cases}\\
\sin^{2}{2\theta_{13}}&=0.093\pm0.008 ~.
\end{align}
The mass-squared differences, which are also important
parameters, are:
\begin{align}
\Delta m^{2}_{21}&=7.53\pm0.18\ (10^{-5}{\rm eV^{2}})\\
\Delta m^{2}_{32}&=
\begin{cases}
2.52\pm0.07\ (10^{-3}{\rm eV^{2}})\ ({\rm normal\ mass\ hierarchy})\\
2.44\pm0.06\ (10^{-3}{\rm eV^{2}})\ ({\rm inverted\ mass\ hierarchy}) ~.
\end{cases}
\end{align}
In this paper, we assume the normal hierarchy scenario
for the neutrino masses, and take the values
\begin{align}
    m_{\nu,1}&=10^{-6}\ {\rm eV}\\
    m_{\nu,2}&=\sqrt{m^{2}_{\nu,1}+\Delta m^{2}_{21}}\cong0.0087\ {\rm eV}\\
    m_{\nu,3}&=\sqrt{m^{2}_{\nu,2}+\Delta m^{2}_{32}}\cong0.050\ {\rm eV}
\end{align}
and, for the mixing angles,  
\begin{align}
    s_{12}=0.55,\ s_{23}=0.66,\ s_{13}=0.15~.
\end{align}
Concerning the complex phases, we take
\begin{align}
\delta=\alpha_{21}=\alpha_{31}=0 ~, \label{eq:tmp_2.19}
\end{align}
for simplicity.  Another free parameters are the
$3\times 3$ elements of $M_N$.  Although it is known
that the structure of this matrix gives an influence 
to the predicted LFV rates~\cite{Casas:2001sr,Blazek:2001zm,
Ibarra:2009bg,Alonso:2011jd,Park:2013msa},
in this paper we assume 
\begin{align}
(M_N)_{ij}=M_{\nu}\times\delta_{ij}~, \label{eq:tmp_2.20}
\end{align}
where $M_\nu$ is a real number.

Under the assumption of Eq.~(\ref{eq:tmp_2.20}), we can determine the
neutrino Yukawa coupling matrix $Y_N$ from Eq.~(\ref{eq:tmp_2.8}) by
using the input data Eqs.~(\ref{eq:tmp_2.10})--(\ref{eq:tmp_2.19}).
How exactly to do this is well known
in literature (see e.g.\ \cite{Hisano:1995cp}), and below is a
brief summary of the procedure.  First, in the basis where the
charged-lepton Yukawa coupling matrix is diagonal, $Y_N$
can be expressed by using two unitary matrices $U$ and $V$
and a diagonal matrix $Y_N^D$ as 
\begin{align}
   (Y_N)_{ij} = V_{ik} (Y_N^D)_{k\ell} U_{\ell j} ~.
\end{align}
When the right-handed neutrino Majorana mass matrix is diagonal,
by a suitable redefinition of the right-handed neutrino
superfields, we can take the matrix $V$ to be a unit matrix
without loss of generality:
\begin{align}
  (Y_N)_{ij} = (Y_N^D)_{i} U_{ij} ~.
\end{align}
where $(Y_N^D)_{i}$ $(i=1,\ldots, 3)$ are the diagonal
entries of the matrix
$(Y_N^D)_{ij}$, namely, $(Y_N^D)_{ij} = (Y_N^D)_{i} \delta_{ij}$.  
By substituting this into Eq.~(\ref{eq:tmp_2.8}), we can identify $U$
as $U_{\text{MNS}}$ and determine $(Y_N^D)_i$ to be
\begin{align}
  (Y_N^D)_i = \sqrt{ \frac{ M_\nu m_{\nu,i} }{ v^2 \sin^2 \beta }} ~,
 \label{eq:tmp_YND}
\end{align}
that is, 
\begin{align}
   (Y_N)_{ij} = \sqrt{ \frac{ M_\nu m_{\nu,i} }{ v^2 \sin^2 \beta }}
                (U_{\text{MNS}})_{ij} ~.
\end{align}
Later in this paper we use this expression to calculate 
$\text{Br}(\mu \to e\gamma)$.

\section{\Large{Lepton Flavor Violation}}
In this section we discuss charged lepton flavor
violation, taking the NMSSM $+~\nu_R$ model as an example
of new physics beyond the SM.

\subsection{$\mu\to e\gamma$ in the Standard Model with $\nu_R$}

Within the SM, the neutrinos are strictly massless and
lepton flavor number is exactly conserved.
The experimental observations of neutrino oscillations~\cite{Agashe:2014kda},
however, make it clear that we have to extend the SM in such a way
that it can accommodate the neutrino masses and mixings.  One of
the simplest extensions is to introduce right-handed neutrinos
($\nu_R$) which are singlet under the SM gauge group, which allows
us to introduce Dirac mass terms for the neutrinos in the Lagrangian.

Once we introduce the right-handed neutrinos in the SM,
in general, charged lepton flavor number is no longer conserved.
This is similar to the case in the quark sector, and the
mismatch between the gauge eigenstates and the mass eigenstates
violates the lepton flavor number conservation.
The branching ratio of $\mu\to e\gamma$ in this model
is given by~\cite{Cheng:1976uq,Petcov:1976ff,Bilenky:1977du}
\begin{align}
 {\rm Br}(\mu\to e\gamma)
 =\frac{\alpha}{32\pi}\left|
 \sum_{i=2,3} (U_{\rm MNS})^*_{ei}(U_{\rm MNS})_{\mu i}
  \frac{m^{2}_{\nu,i}-m^{2}_{\nu,1}}{M^{2}_W}\right|^2 ~.
\end{align}
The suppression factor $(m^2_{\nu,i}-m^2_{\nu,1})^2/M^4_W$
makes the branching ratio extremely small, and it is very
difficult for near future experiments to detect $\mu \to e\gamma$
in this model.  On the contrary, in the non-minimal extensions
of the SM such as the (N)MSSM$+\nu_R$, sizable LFV rates can be 
predicted depending on the parameter region, and this makes
the LFV searches very important as a probe of new physics beyond
the SM.

\subsection{$\mu\to e\gamma$ in NMSSM $+ ~\nu_R$ Model}

%
In the NMSSM$+\nu_R$ model, there are two diagrams which
give dominant contributions to the $l_i \to l_j \gamma$
decays (where $i$ and $j$ are the generation indices which
run from 1 through 3 with $i>j$). 
One is the diagram
with the neutralino and the charged slepton in the loop, and the other
is the diagram which involves the chargino and the sneutrino.
In general, the amplitude $T$ for the
$l_i\to l_j\gamma$ decay can be written as
\begin{align}
T = e m_{l_i} \epsilon^{\alpha*}\bar{u}_j(p-q)
\left[
i\sigma_{\alpha\beta}q^{\beta}\left(A^{L}_2 P_L + A^{R}_2 P_R\right)
\right]u_i(p) ~,
\end{align}
where $e$ is the positron charge, $\epsilon^\alpha$ is the polarization
vector of the photon, $u_i$ and $u_j$ are the spinors for the
initial- and final-state leptons, respectively.  
The momenta $p$ and $q$ are the incoming momentum of the initial
state lepton $l_i$ and the outgoing momentum of the final state
photon, respectively.
The operators $P_{L,R}$ stand for the chiral projection operators.  
The dependence of the amplitude on the models is included
in the coefficients $A^{L}_2$ and $A^{R}_2$.
In the case of the MSSM$+\nu_R$ model, the explicit forms of
$A^{L}_2$ and $A^{R}_2$ are given, for example, in
Refs.~\cite{Borzumati:1986qx,Hisano:1995nq,Hisano:1995cp}.
In the case of the NMSSM $+~\nu_R$ model, they are essentially the
same as the MSSM $+~\nu_R$ model,
except that there are five neutralinos, instead of four,
at low energies, and we can use the expressions in
Refs.~\cite{Borzumati:1986qx,Hisano:1995nq,Hisano:1995cp}
with small modifications.
By using the formulas mentioned above,
the decay branching ratio ${\rm Br}(l_i\to l_j\gamma)$ can be
calculated from the amplitudes to be
\begin{align}
 {\rm Br}(l_i\to l_j\gamma)
 =\frac{e^2}{16\pi}\frac{m^{5}_{l_i}}{\Gamma_{l_i}}
  (|A^{L}_2|^{2}+|A^{R}_2|^{2})~,
\end{align}
where $\Gamma_{l_i}$ is the total decay width of the lepton $l_i$.

A comment on the singlino (i.e., the fermionic component
of the singlet Higgs chiral superfield $\hat{S}$) is in
order here.  The neutralino mass matrix ${\cal M}_{\tilde{\chi}^0}$
in the NMSSM is~\cite{Ellwanger:2009dp}:
\begin{align}
 {\cal M}_{\tilde{\chi}^0}
= \begin{pmatrix} 
M_1 & 0   & - g_1 v_d / \sqrt{2} &   g_1 v_u / \sqrt{2} & 0 \\
0   & M_2 &   g_2 v_d / \sqrt{2} & - g_2 v_u / \sqrt{2} & 0 \\
    &     &   0                  & - \mu_{\text{eff}}   & - \lambda v_u \\
    &     &                      &  0                   & - \lambda v_d \\
\text{(symm.)} &  &              &                      & 2 \kappa s
 \end{pmatrix} ~,
\end{align}
where ``symm.'' means that this matrix is symmetric, 
${\cal M}_{\tilde{\chi}^0}^\top = {\cal M}_{\tilde{\chi}^0}$,
and $\mu_{\text{eff}} \equiv \lambda s$, 
$v_u \equiv v \sin\beta$ and $v_d \equiv v \cos\beta$
where $\beta \equiv \arctan(\langle H_u \rangle / \langle H_d \rangle)$.
The parameters $M_1$ and $M_2$
are the $U(1)_Y$ and $SU(2)_L$ gaugino masses, respectively,
and $g_1$ and $g_2$ the $U(1)_Y$ and $SU(2)_L$ gauge couplings,
respectively.
Since for our sample parameters discussed in Section 5
the value of $\lambda$ is taken to be as small as 0.1, 
and since the value of the $(5,5)$ component of 
${\cal M}_{\tilde{\chi}^0}$ for our sample parameters
is ${\cal O}(1-10) \text{TeV}$, which is much larger
than $\lambda v_u$ and $\lambda v_d$, the mixing between
the singlino and the other components of the neutralinos are
suppressed by the small values of the (3,5), (4,5), (5,3) and (5,4)
entries of ${\cal M}_{\tilde{\chi}^0}$.  
In addition, since the singlino does not couple to the MSSM
matter fields at tree level, the contribution from the singlino
component to the LFV rate is negligible for our sample parameters.

In order to have a non-vanishing LFV rate,
we must have
off-diagonal elements in the slepton mass matrices.
The mass matrices are given as,
\begin{align}
M^2_{\tilde{l}}&=
\left(
\begin{array}{cc}
M^{2}_{LL}&M^{2}_{LR}\\
M^{2}_{RL}&M^{2}_{RR}
\end{array}
\right) ~, \\
\left(M^{2}_{\tilde{\nu}}\right)_{ij}&=m^2_{L,ij}+\frac{1}{2}M^{2}_{Z}\cos{2\beta}\delta_{ij} ~,
\end{align}
where $M^{2}_{LL}, M^{2}_{RR}, M^{2}_{LR}, M^{2}_{RL}$
are the $3\times3$ matrices whose $(i,j)$ elements are given as
\begin{align}
 \left(M^{2}_{LL}\right)_{ij}
 &=m^2_{L,ij}+v^{2}_{d}\left(Y^{\dagger}_{E}Y_E\right)_{ij}
 +M^{2}_{Z}\cos{2\beta}(-\frac{1}{2}+\sin^{2}{\theta_{W}})\delta_{ij}~,\\
 \left(M^{2}_{RR}\right)_{ij}
 &=m^2_{E,ij}+v^{2}_{d}\left(Y^{\dagger}_{E}Y_E\right)_{ij}
 -M_{Z}^{2}\cos{2\beta}\sin^{2}{\theta_{W}}\delta_{ij} ~, \\
 \left(M^{2}_{LR}\right)_{ij}
 &=v_d\left(\left(A^{\ast}_E\right)_{ij}+\mu\tan{\beta}\right)
 \left(Y_E\right)_{ij}~,\\
\left(M^{2}_{RL}\right)_{ij}&=\left((M^{2}_{LR})^{\dagger}\right)_{ij}~.
\end{align}
%
In this paper, we assume mSUGRA-like boundary conditions,
in which all the SUSY breaking parameters that have flavor
indices do not have flavor mixings at the GUT scale.
This means that there are no off-diagonal elements in
the matrices $M^2_{\tilde{l}}$ and $M^2_{\tilde{\nu}}$.  However,
off-diagonal elements in these mass matrices are
induced by radiative corrections
at the energy scale higher than $M_N$, which can be seen in the RGE, 
\begin{align}
16\pi^2 \frac{d}{dt}(m_L)^2_{ij}
&=\left(16\pi^2 \frac{d}{dt}(m_L)^2_{ij}\right)_{\rm {NMSSM}} \nonumber\\
 & \quad\quad
 + (Y_N^\dagger Y_N m_{L}^2)_{ij} + (m_{L}^2 Y_N^\dagger Y_N)_{ij} 
+ 2(Y_N^\dagger (m_{N}^2)^\top Y_N)_{ij}\nonumber\\
 & \quad\quad + 2(Y_N^\dagger Y_N)_{ij}  m_{H_u}^2
+ 2(T_N^\dagger T_N)_{ij} ~,
\end{align}
where $t = \ln{Q}$ with $Q$ being the renormalization scale
and $(T_N)_{ij}$ $(i,j=1,\ldots,3)$ are the trilinear coupling
among the right-handed sneutrino $\tilde{\nu}_R$, the left-handed
slepton $\tilde{L}$, and the Higgs field $H_u$, 
\begin{align}
{\cal L}_{\text{neutrino trilinear}} =  
- (T_N)_{ji} H_{u}\cdot \tilde{L}_{i} \tilde{\nu}^{*}_{Rj} ~ .
\end{align} 
The RGE above directly means that both 
$M^2_{\tilde{l}}$ and $M^{2}_{\tilde{\nu}}$ have off-diagonal elements
at low energies.  The size of these off-diagonal elements can
be roughly estimated as~\cite{Borzumati:1986qx,Hisano:1995nq,Hisano:1995cp},
\begin{align}
\label{eq:logappx}
(\Delta m^{2}_{L})_{ij}=-\frac{1}{16\pi^2}\ln{\frac{M_{{\rm GUT}}}{M_\nu}}(6m^{2}_{0}+2A^{2}_{0})(Y_{N}^{\dagger}Y_{N})_{ij} ~,
\end{align}
where $i\ne j$.  As is clear from Eq.~(\ref{eq:logappx}), 
the slepton off-diagonal elements in this model
comes from the neutrino Yukawa couplings, $Y_N$.
The branching ratio can be estimated in terms of the 
off-diagonal elements to be~\cite{Hisano:1995cp}
\begin{align}
\label{eq:cLFV_apx}
{\rm Br}(l_i\to l_j\gamma)\sim 
\frac{\alpha^3}{G^2_F}\frac{((m^{2}_{L})_{ij})^2}{M_{{\rm SUSY}}^8}~.
\end{align}
At this moment, the most stringent experimental constraint on the 
$\mu\to e\gamma$ is given by the MEG experiment and the 
upper limit is $5.7\times10^{-13}$~\cite{Agashe:2014kda,Adam:2013mnn}.
This bound will be further improved by the upgraded MEG
experiment to $\sim6\times10^{-14}$~\cite{Baldini:2013ke},
and this makes the experiment very important as a probe
of new physics beyond the SM.

\subsection{Other cLFV processes}

In this paper we focus on $\mu\to e\gamma$ in the later sections,
but there are many other charged LFV processes~\cite{Kuno:1999jp}.
Here we mention some of them.

There are two other $l_i\to l_j\gamma$ processes,
$\tau \to \mu \gamma$ and $\tau \to e\gamma$.
Their current experimental limits are
${\rm Br}(\tau\to e\gamma)<3.3\times10^{-8}$ and
${\rm Br}(\tau\to\mu\gamma)<4.4\times10^{-8}$~\cite{Agashe:2014kda}.
In the near future, these limits are expected to be improved to the
level ${\rm Br}(\tau\to l\gamma)<1.0\times10^{-9}$
at Belle-II~\cite{Aushev:2010bq}.  Under the assumptions
we set out at Section 2, the $\mu \to e\gamma$ decay
is more sensitive to SUSY particles, and hence we focus on
$\mu\to e\gamma$ in this paper. 

Other important cLFV processes include
$l^{+}_i\to l^{+}_j l^{+}_jl^{-}_{j}$ and
$\mu\mbox{-}e$\ conversion in nuclei.
As for the former process, when the photon mediation
diagram is dominant, the branching ratio can be
related to that of the $l_i\to l_j\gamma$ decay as~\cite{Hisano:1995cp}
\begin{align}
 {\rm Br}(l^{+}_i\to l^{+}_jl^{+}_jl^{-}_{j})
 \sim \frac{\alpha}{8\pi}
 \left(\frac{16}{3}\ln{\frac{m_{l_i}}{2m_{l_j}}}
 -\frac{14}{9}\right){\rm {Br}}(l_i\to l_j\gamma)~,
\end{align}
and hence ${\rm Br}(l^{+}_i\to l^{+}_jl^{+}_jl^{-}_{j})$
can be calculated once 
${\rm {Br}}(l_i\to l_j\gamma)$ is obtained.
The current experimental limit 
for $\mu^{+}\to e^{+}e^{+}e^{-}$ is
${\rm Br}(\mu^{+}\to e^{+}e^{+}e^{-})<1.0\times
10^{-12}$~\cite{Agashe:2014kda},
and this is expected to be improved to 
${\rm Br}(\mu^{+}\to e^{+}e^{+}e^{-})<1.0\times10^{-16}$
at the Mu3e experiment at PSI~\cite{Mu3e:2012}.
Concerning the $\mu\mbox{-}e$ conversion in nuclei,
there is a simple relation between the conversion
rate $B_{\mu e}(N)$ and ${\rm Br}(\mu\to e\gamma)$
when the photon mediation diagram gives the dominant
contribution~\cite{Kitano:2002mt},
\begin{align}
B_{\mu e}(N)=R(Z){\rm Br}(\mu\to e\gamma)~,
\end{align}
where $B_{\mu e}(N)\equiv
\Gamma(\mu^{-}N\to e^{-}N)/\Gamma(\mu^{-}N\to {\rm capture})$
is the conversion rate normalized to the muon capture
rate $\Gamma(\mu^{-}N\to {\rm capture})$,
and $R(Z)$ is a parameter which depends on the atomic
number $Z$ of the nucleus which captures the muon.
The current limits are $B_{\mu e}({\rm Ti})<4.3\times10^{-12},
B_{\mu e}({\rm Au})<7\times10^{-13}$~\cite{Agashe:2014kda}.
The near future experiments are the COMET experiment
at J-PARC~\cite{COMET:2007}, the Mu2e experiment
at FNAL~\cite{mu2e:2007}, and the PRISM/PRIME experiment at
J-PARC~\cite{PRISM:2006}, which are expected to improve
the bounds to $B_{\mu e}({\rm Al})\sim 6\times10^{-17}$~\cite{Kuno:2013mha},
$B_{\mu e}({\rm Al})\sim 6\times10^{-17}$~\cite{Brown:2014wba},
$B_{\mu e}({\rm Al})\sim 10^{-18}$~\cite{PRISM:2006}, respectively.
Since the $R(Z)$ factors for these experiments are
$R \sim 0.0025$ for $\text{Al}$ 
and $R \sim 0.0040$ for $\text{Ti}$~\cite{Kitano:2002mt},
these experiments are expected to go beyond the corresponding
limit of the $\mu\to e\gamma$ decay by $1.5 \sim 3$ orders
of magnitude, and this will be very useful to probe broader
parameter region of new physics.

\section{\Large{Constraints on the Parameters in the Model}}
In this section we discuss constraints on the parameters
in the NMSSM $+ ~\nu_R$ model.  Some of the issues below are
already discussed in literature~\cite{Ellwanger:2009dp}.

\subsubsection*{Tadpole conditions}

In the NMSSM, there are three tadpole conditions.
At tree-level they read:
\begin{align}
 \vu&\left(\mhu + \mueff^{2}+\lambda^{2}\vd^{2}+\frac{\gi^{2}+\gii^{2}}{4}(\vu^{2}-\vd^{2}) \right)
 -\vd \mueff\Beff=0 ~, \label{eq:tadpole_1}\\
 \vd&\left(\mhd + \mueff^{2}+\lambda^{2}\vu^{2}+\frac{\gi^{2}+\gii^{2}}{4}(\vd^{2}-\vu^{2}) \right)
 -\vu \mueff\Beff=0 ~, \label{eq:tadpole_2}\\
 s&\left(\ms+\kappa A_{\kappa}s + 2\kappa^{2}s^{2}+\lambda^{2}(\vu^{2}+\vd^{2})-2\lambda\kappa\vu\vd\right)
 -\lambda\vu\vd A_{\lambda}=0~, \label{eq:tadpole_3}
\end{align}
where $\mueff=\lambda s$ and $\Beff=A_{\lambda}+\kappa s$.
We can use these relations to determine three parameters
from other parameters.  For example, we can use these
relations to determine $\mueff, \Beff$ and $\ms$ from the
other parameters.  Later we will discuss which parameters we
use as input.

\subsubsection*{Maximal Tree-level Higgs Mass condition}

One of the advantages of the NMSSM over the MSSM is that there is
a parameter region in which the lightest Higgs
boson mass can be made larger than that of the MSSM.
As can be seen from Eq.~(\ref{eq:apx_higgs}),
in order for the Higgs boson mass to be larger,  
it is favorable to have large $\lambda$ and small $\tan\beta$.
The approximate formula Eq.~(\ref{eq:apx_higgs}) is obtained
by neglecting the mixings between the MSSM Higgses and the singlet
Higgs in the CP-even Higgs-boson mass matrix,
\begin{align}
{\cal M}^{2}_{S,{\rm Tree}}=\left( \begin{array}{ccc}
M^{2}_{Z}\cos^{2}{\beta}+\mueff\Beff\tan{\beta}
& (\lambda^{2}v^{2}-\frac{1}{2}M^{2}_{Z})\sin{2\beta}-\mueff\Beff
& \lambda\left(2\mueff v_{d} - \left( \Beff + \kappa s\right)v_{u}\right)\\
& M^{2}_{Z}\sin^{2}{\beta} +\mueff\Beff\cot{\beta} 
& \lambda\left(2\mueff v_{u} - \left( \Beff + \kappa s\right)v_{d}\right)\\
(\text{symm.})&
& \kappa s\left(A_{\kappa}+4\kappa s \right) + \lambda A_{\lambda}\frac{v_{u}v_{d}}{s}
\end{array} \right) ~,
\label{eq:tree_higgs_mtx}
\end{align}
where $v_u \equiv v \sin\beta$ and the lower-left components
are related to the upper-right components by the condition
$({\cal M}^{2}_{S,{\rm Tree}})_{ij}=({\cal M}^{2}_{S,{\rm Tree}})_{ji}$. 
If we take the mixing to the singlet Higgs into account,
the lightest Higgs-boson mass reads~\cite{Ellwanger:2009dp}:
\begin{align}
 m^{2}_{h,{\rm NMSSM}}
   \approx\mz^{2}\cos^2{2\beta}
 + \lambda^{2}v^{2}\sin^{2}{2\beta}
 - \frac{\lambda^{2}}{\kappa^{2}}v^{2}
 \left(\lambda-\left(\kappa+\frac{A_{\lambda}}{2s}\right)
 \sin{2\beta} \right)^2
 + \Delta m^{2}_{h,{\rm 1loop}}
\label{eq:full_Higgs}~.
\end{align}
As can be seen from this equation, the mixing to the
singlet Higgs makes the tree-level lightest
Higgs-boson mass smaller.
The $\lambda$ dependence of the lightest
Higgs-boson mass mainly comes from the second and
third terms, and too large value of $\lambda$ makes
the Higgs boson very small.
There are two ways to decrease the mixing with the
singlet:  One way is to assume a small $\lambda ~(\lesssim 0.1)$,
and the other is to tune the parameters to satisfy
the relation\footnote{The sum of the first three terms
on the right-hand side of Eq.~(\ref{eq:full_Higgs}) is
not an exact expression for the lightest Higgs-boson mass
at tree level, but just an approximation for it.
Therefore, even after making the third term vanish
by imposing the condition Eq.~(\ref{eq:mHM_condi}),
the sum of the first three terms
does not necessarily agree with the exactly maximal value 
of the tree-level lightest Higgs-boson mass obtained by diagonalizing
the full $3\times 3$ Higgs boson mass matrix,
Eq.~(\ref{eq:tree_higgs_mtx}), but only approximately.
However, when the two conditions
$\mueff\Beff\gg v^{2}_u, v^{2}_d$ and
$4\kappa^{2}s^{2}\gg 2\lambda\mueff v-(\Beff+\kappa s)v\sin{2\beta}$
are satisfied, this approximation holds with a very good accuracy.},
\begin{align}
 \lambda - \left( \kappa + \frac{A_{\lambda}}{2s} \right)
 \sin 2\beta =0~.
\label{eq:mHM_condi}
\end{align}

\subsubsection*{Conditions from positive
   CP-even and CP-odd Higgs boson mass-squared}

The $(3,3)$ element in the CP-odd Higgs-boson mass
matrix is given as,
\begin{align}
({\cal M}^{2}_{P})_{33}=4\lambda\kappa\vu\vd + \lambda A_{\lambda}
 \frac{\vu\vd}{s} - 3\kappa s A_{\kappa} ~,
 \label{eq:tmp4.8}
\end{align}
where, in the sample parameter region we study in this paper,
the third term on the right-hand
side gives the dominant contribution.  Therefore,
in order for the CP-odd Higgs mass-squared to be positive,
we must have the condition,
\begin{align}
\kappa s A_{\kappa} \lesssim 0~,
\end{align}
in the approximation that the first and second terms in
Eq.~(\ref{eq:tmp4.8}) are negligible compared to the third term.

Another condition is that the $(3,3)$ element of the
CP-even Higgs-boson mass-squared matrix
\begin{align}
({\cal M}^{2}_{S})_{33}=\lambda A_{\lambda} \frac{\vu\vd}{s} + \kappa s\left(A_{\kappa} + 4\kappa s \right)~,
\label{eq:higgs33}
\end{align}
should be positive:
\begin{align}
-4(\kappa s)^{2} \lesssim \kappa s A_{\kappa}~,
\end{align}
where we have worked in the approximation $s \gg v_u, v_d$.
This condition comes from the requirement that the
singlet Higgs-boson mass-squared must be positive
in the approximation that the mixing between the singlet
and any of the MSSM Higgs doublets is neglected.
Summing up, the condition which $A_\kappa$ should
satisfy is 
\begin{align}
-4(\kappa s)^{2} \lesssim \kappa s A_{\kappa} \lesssim 0~.
\end{align}
In the numerical analysis presented in this paper,
we give $A_{\kappa}$ as an input parameter at the SUSY scale.

\subsubsection*{Constraint from non-vanishing VEV of $S$}

There is a condition on the model parameters from
the requirement that
the singlet Higgs $S$ has a non-zero VEV, $\langle S \rangle
\equiv s \neq 0$.
When $s\gg\vu, \vd$, the potential for $S$ reads:
\begin{align}
 V(S)\sim m^{2}_{S}S^{2}
  +\frac{2}{3}\kappa A_{\kappa}S^{3}+\kappa^{2}S^{4}~.
\end{align}
If we require that this potential has a minimum
at $S=s \neq 0$, and that the value of $V(S)$ at $S=s$ is
smaller than $V(0)$, we obtain the condition~\cite{Ellwanger:2009dp},
\begin{align}
  A^{2}_{\kappa} \gtrsim 9m^{2}_{S}~.
\end{align}

\subsubsection*{Constraint from Perturbativity of $\lambda$}

The tree-level Higgs boson mass becomes larger for larger
value of $\lambda$ unless we take the mixing with the singlet
into account.  However, there is a limit on the size of
$\lambda$ which comes from theoretical consideration.
Namely, in order for $\lambda$ not to blow up below
the GUT scale, the value of $\lambda$ at the SUSY scale
must be smaller than $\sim 0.7$~\cite{Ellwanger:2009dp}.

\subsubsection*{Condition from the SM-like lightest Higgs boson}

In this paper, we identify the lightest CP-even Higgs boson
as the Higgs boson discovered at the LHC~\cite{Agashe:2014kda}.
The properties of the discovered particle such as the
decay branching ratios are known to be consistent with
those of the Higgs boson in the minimal SM.
This means that we have to require that the lightest CP-even
Higgs boson in the model we consider 
should not be singlet-like but like the
lightest Higgs boson in the MSSM which is known to become
SM-like in the decoupling limit.

\section{\Large{Numerical Results}}
In this section, we give our numerical results.

First, we explain how we choose independent input parameters.
To maximally keep the similarity to the cMSSM, we choose
$\tan\beta$ at the SUSY scale and $m_0$, $M_{1/2}$ and $A_0$
at the GUT scale as input parameters.  In addition, since the
parameter $\lambda$ directly enters in the expression for the
lightest Higgs-boson mass, we choose $\lambda$ at the SUSY scale
as input.  If we further choose either $\kappa$ or $A_\lambda$
as input, we can use the two tadpole
conditions Eqs.~(\ref{eq:tadpole_1}) and (\ref{eq:tadpole_2})
to determine $\mueff (=\lambda s)$ and
$\Beff (=\kappa s + A_{\lambda})$,
and then use Eq.~(\ref{eq:tadpole_3}) to fix $m^2_S$
by using the value of $A_\kappa$ as an additional input.
Below we consider two cases: in one case we choose $\kappa$
at the SUSY scale as input, and in the other case we take
$A_{\lambda}$ at the GUT scale as input.
Summing up, we consider two sets of input parameters.
In one case, we choose the parameters below as input,
\begin{alignat}{3}
&  \tan{\beta}~,~ \lambda~,~ \kappa~,~ A_{\kappa}~ & ~~~
 &\text{at the SUSY scale}~,
 \nonumber \\
& m_{0}~,~ M_{1/2}~,~ A_{0}~  &~~~&\text{at the GUT scale} ~,
\end{alignat}
which we call the case 1, and in the other case, 
we take the parameters below as input:
\begin{alignat}{3}
 & \tan{\beta}~,~ \lambda~,~  A_{\kappa}~& ~~~&
 \text{at the SUSY scale} ~,\nonumber \\
& m_{0}~,~ M_{1/2}~,~ A_{0}~,~ A_{\lambda}~&
 ~~~&\text{at the GUT scale} ~,
\end{alignat}
which we call the case 2.

\subsection*{\underline{Case 1}}

In this case, we determine the parameters
$s = \mueff/\lambda$ and $A_{\lambda}=\Beff-\kappa s$
by using the tadpole conditions.
If we are to use Eq.~(\ref{eq:mHM_condi}),
we have to tune $\kappa$ to satisfy Eq.~(\ref{eq:mHM_condi}).
The value of $\kappa$ in this case is
\begin{align}
\kappa = \frac{\lambda}{\sin{2\beta}} - \frac{A_{\lambda}}{2 s}~.
\label{eq:tmp_5.7}
\end{align}
This equation means that for large $\tan\beta$ and for large
$\lambda$, the $\kappa$ parameter becomes too large, and then
$\lambda$ at the scale higher than the weak scale becomes
too large to be perturbative, and eventually
it blows up below the GUT scale\footnote{
For $\lambda=0.3$ and $\tan{\beta}=3$,
$\kappa$ is approximately $\kappa\sim0.5$ (the second
term of Eq.~(\ref{eq:tmp_5.7}) is small (typically
${\cal O}(0.1)$ or less) for large part of our sample parameters).
The RGEs for $\lambda$ and $\kappa$ are
\begin{align}
16\pi^2 \frac{d}{dt} \lambda &= \lambda
\left[ 2|\kappa|^2 + 4|\lambda|^2
    + 3 \Tr(Y_U^\dagger Y_U) + 3 \Tr(Y_D^\dagger Y_D)
    +   \Tr(Y_E^\dagger Y_E) +   \Tr(Y_N^\dagger Y_N)
    - 3 g_2^2 - \frac35 g_1^2 \right]~, \\
16\pi^2 \frac{d}{dt} \kappa &= \kappa
\bigg[   6|\kappa|^2 + 6|\lambda|^2 \bigg]~.
\end{align}
If we assume (\ref{eq:mHM_condi}), then a large
$\lambda$ induces a large $\kappa$ via RGEs, and
$\lambda$ can develop the Landau pole below the GUT scale
depending on the parameters.
For small $\tan{\beta}$, the large top Yukawa coupling
makes the right-hand side of the RGE for $\lambda$ large,
and this makes it easier for the Landau pole for $\lambda$
to occur.}.
We therefore do NOT assume Eq.~(\ref{eq:mHM_condi}) for the case
1, and assume small $\lambda~(\sim 0.1)$ to make
the mixing of the MSSM Higgses with the singlet Higgs smaller,
in order not to decrease the tree-level Higgs-boson mass.

\begin{figure}[t]
 \begin{center}
  \subfigure[$\kappa=0.09$]{
   \includegraphics[width=.50\columnwidth]{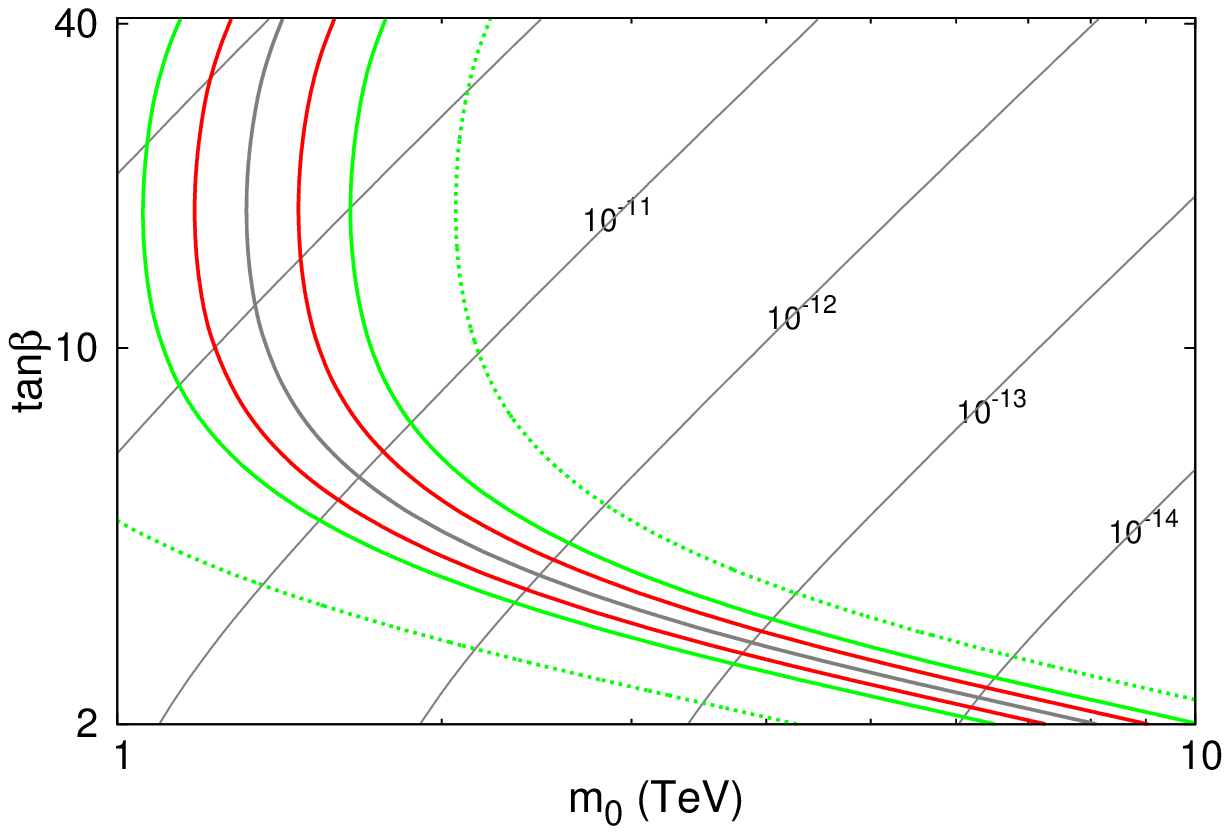}
  }~
  \subfigure[$\kappa=0.05$]{
   \includegraphics[width=.50\columnwidth]{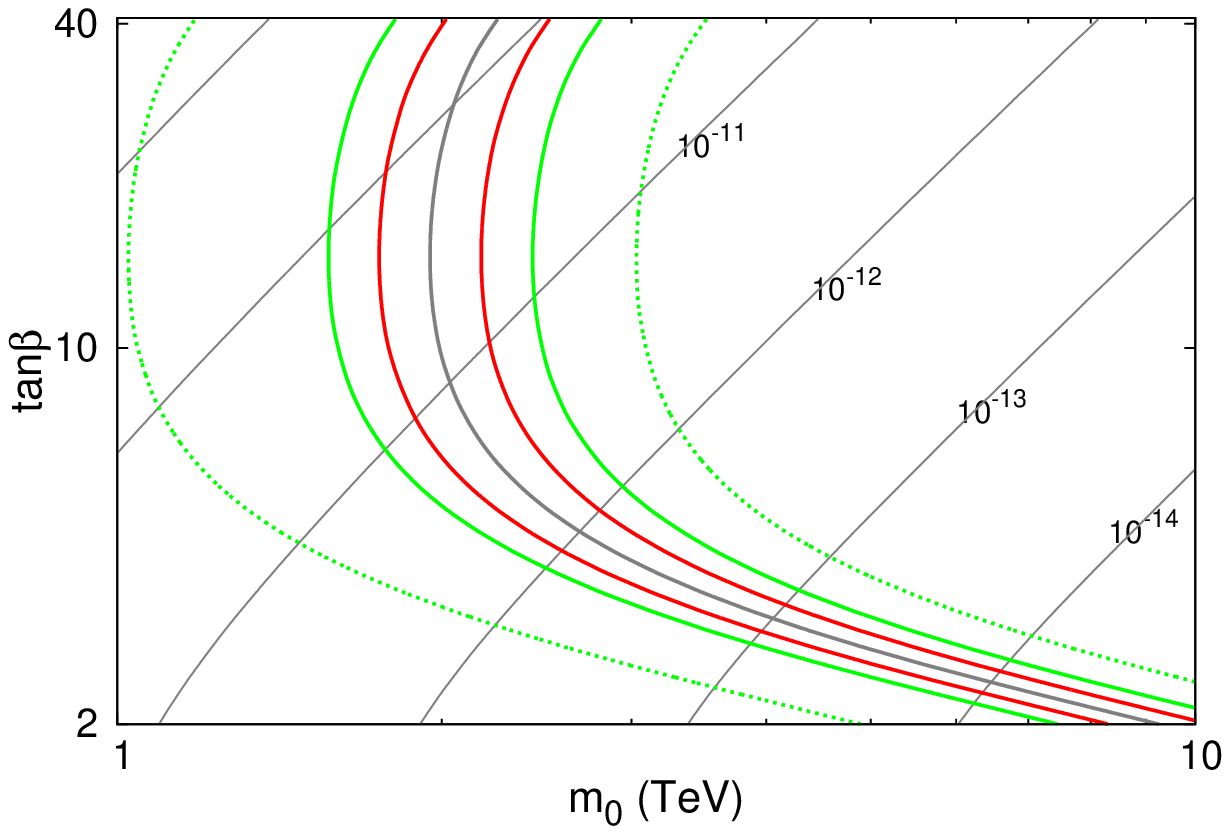}
  }
  \caption{\label{fig:T2P2} Our numerical results on
  ${\rm Br}(\mu \to e\gamma)$ and the lightest Higgs boson mass
  in the MSSM-like semi-constrained NMSSM, case 1.
  The diagonal gray contours are the contours for the predicted branching
  ratios of $\mu\to e\gamma$.  The diagonal solid and dotted red
  lines are the current limit and the near-future expected sensitivity
  of ${\rm Br}(\mu\to e\gamma)$, respectively.
  The current limits of the $\mu \to 3e$ and $\mu$-$e$ conversion
  rates are also shown by the corresponding values of
  ${\rm Br}(\mu \to e \gamma)$ as the solid light-blue
  and dark-blue lines, respectively.  The near-future expected reach
  of ${\rm Br}(\mu \to 3e)$ is shown by the dotted light-blue lines,
  and that of the $\mu$-$e$ conversion at COMET and Mu2e 
  is shown by the dotted
  dark-blue lines.   The gray line is the contour
  for the 126 GeV Higgs boson mass.  The regions between the
  two red curves, the two solid green curves and the two dotted
  green curves are the areas where the Higgs boson mass is
  in the ranges, 125-127, 124-128 and 120-130 ${\rm GeV}$, respectively.
  We assume $m_0=M_{1/2}$ in the figures.  
  The input parameters at the SUSY scale are
  $\lambda=0.1,A_{\kappa}=-50~{\rm (GeV)}$ and we take
  $A_0=-500~{\rm (GeV)}$ at the GUT scale.
  The right-handed neutrino Majorana masses are taken to be
  $M_{\nu}=5.0\times10^{14}~{\rm (GeV)}$.}
 \end{center}
\end{figure}

\subsubsection*{Numerical Results}

Our numerical results for 
${\rm Br}(\mu \to e\gamma)$ and the Higgs boson mass
in the case 1 are given in Figs.~\ref{fig:T2P2} (a) and (b).
In the figures (a) and (b), $\kappa$ at the
SUSY scale is taken to be 0.09 and 0.05, respectively.
The rest of the input parameters are taken to be the
same in the two figures, and the input SUSY parameters are
$\lambda=0.1, A_{\kappa}=-50~{\rm (GeV)}$
at the SUSY scale and $A_0=-500~{\rm (GeV)}$ at the GUT scale.
We take $m_0 = M_{1/2}$, and  
the right-handed neutrino Majorana mass is taken to be
$M_{\nu}=5.0\times10^{14}~{\rm (GeV)}$.
The reason for the choice of this value of $M_\nu$ is that
for $M_{\nu}=5.0\times10^{14}~{\rm (GeV)}$,
the largest neutrino Yukawa coupling becomes ${\cal O}(1)$,
as can be seen from Eq.~(\ref{eq:tmp_YND}).  For smaller
values of $M_{\nu}$, the LFV rates become smaller
since the off-diagonal entries of the slepton mass matrix 
become smaller, see Eq.~(\ref{eq:logappx}).

In the figures, we plot contours for constant values of
${\rm Br}(\mu \to e \gamma)$.  Since ${\rm Br}(\mu \to e \gamma)$
is roughly proportional to $\tan^2\beta/m_0^4$ for $A_0 \ll m_0$
and $\tan\beta \gg 1$, the dependence of the contours
on $\tan\beta$ and $m_0$ are simple.  In the case of the
MSSM $+\nu_R$ model, similar results are known
in literature~\cite{Ibarra:2009bg, Alonso:2011jd, Park:2013msa,
Borzumati:1986qx, Hisano:1995nq, Hisano:1995cp, LFV_MSSMRN}.  

In the figures, we also show the current limit and the 
near-future expected sensitivity of ${\rm Br}(\mu \to e \gamma)$ by
the solid and dotted diagonal red lines, respectively.  
The current limits of the $\mu \to 3e$ and
$\mu$-$e$ conversion rates are also shown by the corresponding
values of ${\rm Br}(\mu \to e \gamma)$ in the figures as
the solid light-blue and dark-blue lines, 
respectively.  Similarly, the near-future expected reach
of ${\rm Br}(\mu \to 3e)$ is shown by the dotted light-blue lines,
and that of the $\mu$-$e$ conversion at the COMET and Mu2e
experiments is shown by the dotted dark-blue lines.  
Once the PRISM/PRIME experiment is realized,
it is expected to go beyond the COMET/Mu2e sensitivity by
about two orders of magnitude, and the full region in the
figures will be covered for this particular choice of 
the input parameters. 

Also shown in Figs.~\ref{fig:T2P2} (a) and (b) are
the contours for the lightest Higgs boson mass. 
From the figures, we find that smaller $\kappa$
makes the Higgs boson mass smaller.
We have numerically confirmed that the difference in the Higgs
boson mass mainly comes from the values of $\kappa$, and the
difference in the values of the other parameters like $A_\lambda$
are not very important for the difference in the predictions for
the Higgs boson mass. 
This dependence of the Higgs boson mass on $\kappa$ can be
understood from Eq.~(\ref{eq:full_Higgs}).  Namely,
large $\kappa$ makes the $(3,3)$ element of ${\cal M}^{2}_{S,{\rm
Tree}}$ larger and the mixing between the MSSM Higgses and the
singlet Higgs, which makes a negative contribution to the lightest
Higgs boson mass, smaller.

From the figures, we find that there is a parameter region
which is favored from the Higgs boson mass measurement
where the predicted value of ${\rm Br}(\mu \to e \gamma)$ is
within reach of the near-future experiment even if $m_0$ is
as large as $\sim 4$ TeV.  In addition, the near-future
experiments Mu3e, COMET and Mu2e can probe the SUSY mass
scale up to $\sim 5$ TeV for our sample
parameters.  The reach will be extended further if the PRISM/PRIME
experiment is carried out.

We here comment on the dependence of the Higgs boson mass
on $\kappa$.   In the figures, we take $\kappa$ only down
to 0.05.  For smaller values of $\kappa$, for example, 
$\kappa \lesssim 0.03$ for $\lambda=0.1$, the Higgs boson
mass sharply decreases for decreasing $\kappa$.
This sharp $\kappa$ dependence comes from the factor
$(\lambda/\kappa)^{2}$ in the third term of the right-hand side of
Eq.~(\ref{eq:full_Higgs}).  If we take smaller value of
$\lambda$, this sharp decrease of $\kappa$ happens
at smaller value of $\kappa$, and hence we can take
smaller $\kappa$ as well.

\begin{figure}[t]
 \begin{center}
   \includegraphics[width=.50\columnwidth]{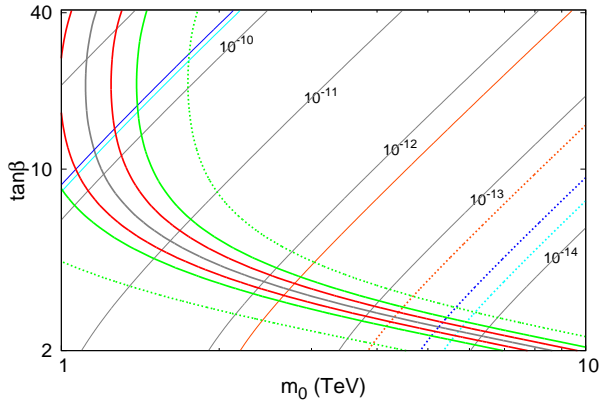}
  \caption{\label{fig:MSSM} The numerical results for
  ${\rm Br}(\mu \to e\gamma)$ and the lightest Higgs boson mass
  in the MSSM $+ \nu_R$ model.
  The diagonal gray contours are the contours for the predicted branching
  ratios of $\mu\to e\gamma$.  
  The meanings of the diagonal red/light-blue/dark-blue 
  solid/dotted lines, gray line, red curves, two solid green curves, and
  two dotted green curves are the same as in Fig.~\ref{fig:T2P2}.
  We assume $m_0=M_{1/2}$ in the figure.  
  We take $A_0=-500~{\rm (GeV)}$ at the GUT scale and assume $\mu > 0$.
  The right-handed neutrino Majorana masses are taken to be
  $M_{\nu}=5.0\times10^{14}~{\rm (GeV)}$.}
 \end{center}
\end{figure}

The difference between the above results and the result
in the case of the MSSM $+\nu_R$ model can become clearer
if we compare Figs.~\ref{fig:T2P2} (a) and (b) with the
prediction in the  MSSM $+\nu_R$ model for similar input
parameters.  In Fig.~\ref{fig:MSSM} we show the prediction
for ${\rm Br}(\mu \to e\gamma)$ in the MSSM $+\nu_R$ model
with the boundary conditions at the GUT scale, 
Eqs.~(\ref{eq:tmp2.3})--(\ref{eq:tmp2.6}) together with
the conditions at the GUT scale  
$M_{1/2}=m_0$ and $(m_{N}^2)_{ij}=m_0^2 \delta_{ij}$, 
where $(m_{N}^2)_{ij}$ $(i,j = 1, \ldots, 3)$ is the soft
SUSY breaking mass-squared matrix for the right-handed neutrinos.
We take $A_0=-500~{\rm (GeV)}$ at the GUT scale
and assume $\mu > 0$.  We also take the same neutrino 
masses and mixing parameters as in Figs.~\ref{fig:T2P2} (a)
and (b), as well as the same right-handed neutrino Majorana masses
which are taken to be $M_{\nu}=5.0\times10^{14}~{\rm (GeV)}$.
By comparing Figs.~\ref{fig:T2P2} (a), (b) with
Fig.~\ref{fig:MSSM}, we see little difference for the 
prediction for ${\rm Br}(\mu \to e\gamma)$ for given 
values of $\tan\beta$ and $m_0$ $(=M_{1/2})$, but
the region favored from the Higgs boson mass slightly
changes.  For example, for a fixed value of $\tan\beta=10$,
the value of ${\rm Br}(\mu \to e\gamma)$ predicted from
the fixed value of the Higgs boson mass at $m_H=126$ GeV
in the MSSM $+ \nu_R$
model is $\sim 10^{-10}$ for this particular choice
of input parameters, while in the case of Figs.~\ref{fig:T2P2}
(a) and (b), the corresponding values are $\sim 6 \times 10^{-11}$
and $\sim 2 \times 10^{-11}$, respectively.  We therefore
conclude that in the NMSSM $+\nu_R$ model, the predicted
value of ${\rm Br}(\mu \to e \gamma)$ favored from the Higgs
boson mass can slightly change from the MSSM $+\nu_R$ model.

\subsection*{\underline{Case 2}}

\begin{figure}[t]
 \begin{center}
  \subfigure[$A_{\lambda}(M_{{\rm GUT}})=-5000~({\rm GeV})$]{
   \includegraphics[width=.50\columnwidth]{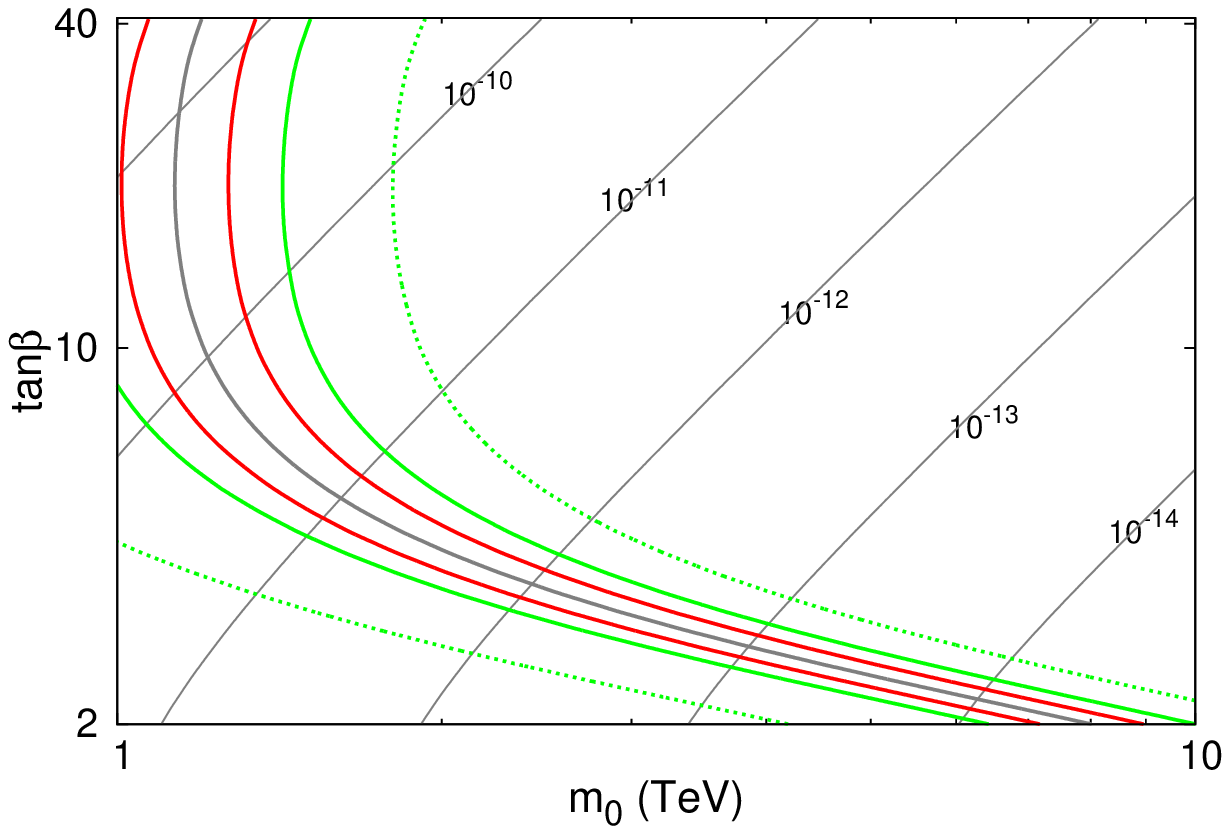}
  }~
  \subfigure[$A_{\lambda}(M_{{\rm GUT}})=-2500~({\rm GeV})$]{
   \includegraphics[width=.50\columnwidth]{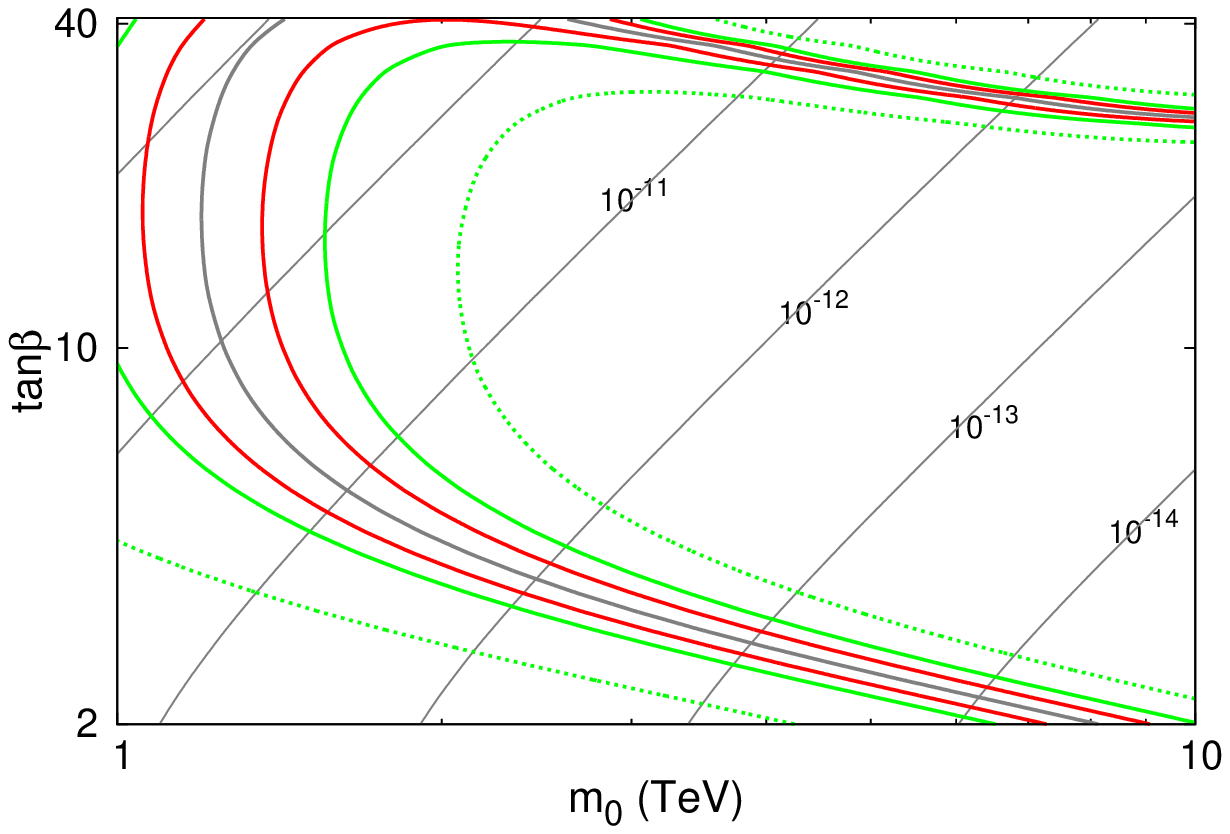}
  }
  \caption{\label{fig:T1P2}  Our numerical results on
  ${\rm Br}(\mu \to e\gamma)$ and the lightest Higgs boson mass
  in the MSSM-like semi-constrained NMSSM, case 2.
  The diagonal gray contours are the contours for 
  the predicted branching ratios of $\mu\to e\gamma$.
  The meanings of the diagonal red/light-blue/dark-blue 
  solid/dotted lines, gray line, red curves, two solid green curves, and
  two dotted green curves are the same as in Fig.~\ref{fig:T2P2}.
  We assume $m_0=M_{1/2}$ in the figures.  
  The input parameters at the SUSY scale are
  $\lambda=0.1, A_{\kappa}=-50~{\rm (GeV)}$ and we take
  $A_0=-500~{\rm (GeV)}$ at the GUT scale.
  The right-handed neutrino Majorana masses are taken to be
  $M_{\nu}=5.0\times10^{14}~{\rm (GeV)}$.}
 \end{center}
\end{figure}

In this case, if we are to use Eq.~(\ref{eq:mHM_condi}), the value of
$\kappa$ is determined to be, similarly to the case 1, 
\begin{align}
\kappa = \frac{\lambda}{\sin{2\beta}} - \frac{A_{\lambda}}{2 s}~.
\end{align}
Similarly to the reasoning in the case 1, the equation
above implies that
if $\tan\beta$ or $\lambda$ is too large,
the $\kappa$ parameter at higher scale blows up and
becomes non-perturbative. 
Therefore, if we are to use Eq.~(\ref{eq:mHM_condi}),
we need small $\lambda$ and small $\tan{\beta}$, but this choice
makes the Higgs boson mass very similar to the
MSSM case and hence is not very interesting.  We therefore do NOT
use Eq.~(\ref{eq:mHM_condi}) in the case 2, either.

\subsubsection*{Numerical Results}

In Figs.~\ref{fig:T1P2} (a) and (b), we give our numerical results
for ${\rm Br}(\mu \to e\gamma)$ and the lightest Higgs boson mass
in the case 2.
In the figures (a) and (b), $A_\lambda$ at the
GUT scale is taken to be $-5000$~GeV and $-2500$~GeV, respectively.
The rest of the input parameters are taken to be the
same in the two figures, and the input SUSY parameters are
$\lambda=0.1,A_{\kappa}=-50 {\rm ~(GeV)}$ at the SUSY scale,
and $A_0=-500 {\rm ~(GeV)}$ at the GUT scale.
We take $m_0 = M_{1/2}$, and $M_{\nu}=5.0\times10^{14}~{\rm (GeV)}$.

In the figures, we plot contours for the constant values of
${\rm Br}(\mu \to e\gamma)$.  The behaviors are very similar
to the case 1.  Also shown are the favored regions
from the Higgs boson mass and the current and near-future
expected sensitivities of the $\mu \to e\gamma$, 
$\mu \to 3e$ and $\mu$-$e$ conversion rates,
similarly to Figs.~\ref{fig:T2P2} (a) and (b).  

From the figures, we find that only in Fig.~\ref{fig:T1P2} (b),
there is an extra Higgs-mass favored parameter space
at the region where $\tan\beta$ and $m_0~(=M_{1/2})$ are both
large.  This difference between the two figures
mainly comes from the difference in the value of $\kappa$, and
the differences in the other parameters like $A_\lambda$
enter only indirectly through the value of $\kappa$
in the prediction for the Higgs boson mass.

We now explain why the changes in the input value of
$A_\lambda$ at the GUT scale affect the value of $\kappa$
at the SUSY scale.

To do so, we first explain the dependence of $\kappa$ on
$m_0~(=M_{1/2})$ and $\tan\beta$ with fixed value of
$A_\lambda(M_{\rm GUT})$.  Below we will show that $\kappa$
becomes smaller for larger $m_0$ and for larger
$\tan\beta$ at the region $\tan\beta \gg 1$
in the parameter space shown in
Figs.~\ref{fig:T1P2} (a) and (b).  At the upper-right
region of Fig.~\ref{fig:T1P2} (b), the value of $\kappa$
becomes $\kappa \lesssim 0.03$, where the Higgs boson mass
decreases relatively quickly for decreasing $\kappa$, as discussed
at the end of the discussion for the case 1 in this
section\footnote{If $\kappa$ is too small, the (3,3) element of the
Higgs boson mass matrix becomes very small and the lightest
Higgs boson becomes singlet-like.  In all the parameter
regions we consider in this paper, the lightest Higgs boson
is MSSM-like.}.
In the parameter region shown in Fig.~\ref{fig:T1P2} (a),
the value of $\kappa$ is
larger than 0.03, and hence this relatively fast
decrease does not happen.
Then we have to explain why $\kappa$ is smaller in
Fig.~\ref{fig:T1P2} (b).  This is because $A_\lambda(M_{\rm SUSY})$
is larger in Fig.~\ref{fig:T1P2} (b) since
$A_\lambda(M_{\rm GUT})$ is larger.   The relation between
$A_\lambda(M_{\rm SUSY})$ and $\kappa$ are given by
$\kappa = (B_{\rm eff} - A_\lambda)/s$ and hence larger
$A_\lambda$ means smaller $\kappa$. 

Let us discuss the change in $\kappa$ for different
$m_0~(=M_{1/2})$ for fixed values of $\tan\beta$
and $A_\lambda(M_{\rm GUT})$.
The value of $A_\lambda$ at the SUSY scale is given by
solving the RGE,
\begin{align}
 16\pi^2 \frac{d}{dt} A_\lambda
&= 
 4|\kappa|^2 A_\kappa + 8 |\lambda|^2 A_\lambda
+ 6 \Tr(Y_U^\dagger T_U) + 6 \Tr(Y_D^\dagger T_D)
+ 2 \Tr(Y_E^\dagger T_E) + 2 \Tr(Y_N^\dagger T_N) \nonumber\\
& \quad + 6 g_2^2 M_2 + \frac65 g_1^2 M_1~.
\label{eq:RGE_alam}
\end{align}
For our sample parameters, $A_{\lambda}(M_{{\rm SUSY}})$
becomes larger\footnote{At first
glance, it appears that larger $m_0~(=M_{1/2})$
makes the gaugino masses larger, which makes the right-hand side of
Eq.~(\ref{eq:RGE_alam}) becomes positive, and that
$A_\lambda$ at the SUSY scale becomes smaller.  However, in
reality, the contributions from $A_t$ and $A_b$ makes
negative contributions, and the balance between the
gaugino mass contributions and the $A_t$ and $A_b$ contributions
determines the scale dependence of $A_{\lambda}$.}
for larger $m_0~(=M_{1/2})$ and fixed $\tan\beta$.
Therefore, for a fixed value of $\tan{\beta}$,
larger $m_0~(=M_{1/2})$ makes $\kappa$ smaller
through the relation, $\kappa = (\Beff - A_{\lambda})/s$.

Next, we discuss the dependence of $\kappa$ on $\tan\beta$,
fixing the values
of $m_0~(=M_{1/2})$ and $A_\lambda$ at the GUT scale.
Since here we are mainly interested in the difference
at large $\tan\beta$ region, in this paragraph we assume
$\tan\beta \gg 1$.
For large $\tan\beta$, $A_\lambda(M_{\rm SUSY})$ becomes
larger for larger $\tan\beta$ since the fourth term
of the right-hand side of Eq.~(\ref{eq:RGE_alam}),
which involves the bottom Yukawa coupling, becomes
more important.  This increase in $A_\lambda(M_{\rm SUSY})$
for larger $\tan\beta$
makes $\kappa$ smaller for fixed $m_0$ since
$\kappa = (\Beff - A_{\lambda})/s$.  Another
reason which makes $\kappa$ smaller for larger $\tan\beta$
comes from the values of $\mueff$ and $\Beff$,
although this effect is less important for large $\tan\beta$.
The values of $\mueff$ and $\Beff$ at the SUSY scale
are obtained by solving the tadpole conditions, and
the solutions at the tree-level are,
\begin{align}
\mu_{{\rm eff}}^2 &=
 \frac{m^{2}_{H_d}-m^{2}_{H_u}\tan^{2}{\beta}}{\tan^{2}{\beta}-1}-\frac{1}{2}M^{2}_{Z}~, \\
B_{{\rm eff}}&= \frac{1}{2\mu_{{\rm eff}}}(m^{2}_{H_u} + m^{2}_{H_d}  + 2 \mu^{2}_{{\rm eff}} + v^{2}\lambda^{2})\sin{2\beta}~.
\end{align}
Both $\mueff$ and $\Beff$ become smaller for larger
$\tan{\beta}$ for our sample parameters.
From the relation $\mueff=\lambda s$,
a smaller $\mueff$ means a smaller $s$ for
fixed $\lambda$.  From $\kappa = (\Beff - A_{\lambda})/s$,
the variation of $\kappa$ comes from that of $s ~(= \mueff/\lambda)$
and that of $\Beff$.  For our sample parameters, since the decrease
in $\Beff$ due to increase in $\tan\beta$
has a larger effect on $\kappa$ than that of $s$, $\kappa$ becomes
smaller for larger $\tan\beta$.

As for ${\rm Br}(\mu \to e \gamma)$, also in the case 2,
we find that there is a parameter
region which is favored from the Higgs boson mass
and in which the predicted value of ${\rm Br}(\mu \to e \gamma)$
is within reach of near-future experiment even if
$m_0 \sim 4 ~({\rm TeV})$, which has not yet been probed at the LHC.

\section{\Large{Summary and Discussions}}
In this paper, we have studied the cLFV in the semi-constrained
NMSSM+$\nu_R$ model, taking into account the recent results on
the Higgs boson mass determination.  We have considered the 
boundary conditions at the GUT scale to be MSSM-like
and semi-constrained in the sense that the SUSY breaking
parameters $A_\lambda, A_\kappa, m_S^2$ which are
specific to the NMSSM are not necessarily equal to
$A_0, A_0, m_0^2$, respectively.  We have considered two
cases: in one case the parameters
($s,\ A_{\lambda},\ m^{2}_{S}$) are determined from the tadpole
conditions, which we call the case 1,
while in the other case ($s,\ \kappa,\ m^{2}_{S}$)
are determined from other input
parameters, which we call the case 2.

One of the advantages of the NMSSM is that the tree-level
lightest Higgs boson mass can be taken to be larger
than that of the MSSM by taking a large value of $\lambda$.
In addition to this effect, there
is another new effect in the Higgs boson sector of the NMSSM,
namely, we also have to take into account the
mixing with the singlet Higgs.  This mixing can decrease
the Higgs boson mass depending on the parameters.
In the semi-constrained scenario we have considered,
we find it is difficult to realize both large $\lambda$
and small mixing with the singlet at the same time.
Hence in this paper we have assumed a small $\lambda ~(\sim0.1)$
which makes the mixing with the singlet small.

In the case 1, we have obtained the results similar to those in
the MSSM $+~\nu_R$ model.  We have also shown that the
Higgs-boson-mass favored
parameter region depends on the value of $\kappa$.  
As the case 2, we have considered the case where the $\kappa$ parameter
is not an input parameter but is a parameter determined
from other parameters via the tadpole conditions, and we
have obtained a partly different favored region from the case 1.
In both cases, we have shown that in the NMSSM+$\nu_R$ model
there is a parameter region
in which the predicted value of ${\rm Br}(\mu\to e\gamma)$
is so large that the $\mu\to e\gamma$ decay can be observable
at the near-future experiment even if the SUSY mass scale
is about 4 TeV.  The reach will be improved further by
the near-future experiments, Mu3e, COMET, Mu2e and PRISM/PRIME.

Several comments are in order.
In this paper we have taken the input SUSY mass parameter
$m_0 (= M_{1/2})$ as high as $\gtrsim$ 1 TeV.  This choice
makes the squarks and gluino as heavy as multi-TeV,
whose possibility is still not excluded by any experiments
including the LHC~\cite{Agashe:2014kda}.  The price
we have to pay is that it is difficult to explain the
muon $g-2$ anomaly in terms of SUSY if we take the multi-TeV
SUSY particle mass scenario, and that we have to
introduce the so-called ``little hierarchy'' between
the weak scale and the SUSY scale.  In particular, some
fine-tuning is necessary in order to keep the Higgs boson
mass protected from large radiative corrections.
Nevertheless, in SUSY models the cancellation between the
bosonic and fermionic loop contributions to the Higgs-boson
mass-squared is automatic at the scales much higher than the
SUSY breaking scale, which decreases the degree of fine-tuning
significantly compared to the non-SUSY minimal standard model
and makes SUSY models still attractive.

\section*{\Large{Acknowledgements}}
K.~N.\  is supported by Tohoku University
Institute for International Advanced Research and Education.
D.~N.\  is a Yukawa Fellow, and this work was partially supported
by the Yukawa Memorial Foundation.
Part of the calculations were carried out
by using the Sushiki computer at the Yukawa Institute for Theoretical
Physics, Kyoto University at the early stages of this work.


\end{document}